# Optical study of chemotherapy efficiency in cancer treatment via intracellular structural disorder analysis using partial wave spectroscopy

Huda M. Almabadi,[a,c] Prashanth K. B. Nagesh,[b] Peeyush Sahay,[a] Shiva Bhandari,[a] Eugene C. Eckstein,[c] Meena Jaggi,[b] Subhash Chauhan,[b] Murali Yallappu,[b] Prabhakar Pradhan [a,*]

[a] The University of Memphis, BioNanoPhotonics Laboratory, Department of Physics and Materials Science, USA, TN38152
[b] University of Tennessee Health Science Center, College of Pharmacy, Department of Pharmaceutical Sciences and the Center for Cancer Research, Memphis, USA, TN 38163
[c] The University of Memphis, Biomedical Engineering, Memphis, USA, TN38152

* *Prabhakar Pradhan, E-mail: ppradhan@memphis.edu*

**Abstract**: With the progress of cancer, the macromolecules such as DNA, RNA, lipids, etc. inside cells undergo spatial structural rearrangements and alterations. Mesoscopic physics based optical partial wave spectroscopy (PWS) was recently introduced to quantify changes in the nanoscale structural disorder in biological cells. The measurement is done in a parameter termed as 'disorder strength' ($L_d$) which represents the degree of nanoscale structural disorder inside the cells. It was shown that cancerous cells have higher disorder strength than the normal cells. In this work, using the PWS technique, we analyzed the hierarchy of different types of prostate cancer cells by quantifying their average disorder strengths. The results showed that increase in the tumorigenicity level of prostate cancer is correlated with the increase in the disorder strength $L_d$. Furthermore, using the $L_d$ parameter we also analyzed docetaxel chemo-resistance attribute of these prostate cancer cells, in comparison to docetaxel chemo-sensitive cancer cells. Our results show that chemo drug resistive cancer cells have mutations to more aggressive, or increased in $L_d$ values relative to the chemo sensitive cancer cells. Potential applications of the $L_d$ parameter for determining the effectiveness of the chemotherapy drug on cancer cells are discussed.

## 1 Introduction

*1.1 Role of nanoscale mass density fluctuations in detection of cancer*

It is now known that the progress of cancer is associated with the nanoscale intracellular structural alteration is cells/tissues.[1,2] Based on an interdisciplinary approach combining mesoscopic physics and biophotonics, optical partial wave spectroscopic microscopy (PWS) was introduced earlier to quantify the nanoscale structural disorder in weakly disordered optical media, such as biological cells.[3,4] PWS approach has shown efficiency in cancer diagnostics [5-7]. From then, there are different versions of the PWS has been developed, with the same goal of measuring nanoscale structural alteration in cells[8,9]. In this work, we have used a newly developed PWS instrument featuring finer focus that provides an accurate scattering volume elements. PWS analysis is based on statistical quantification of the backscattered spectral or back reflected intensities, and their spectral correlation decay length due to the nanoscale refractive index fluctuation within biological samples. Finer focus for reflection volume allows accurate reflection of focus depth vertically, and for compute optical disorder. In PWS technique, a thin sample can be divided into several parallel scattering samples, or the bulk backscattering problem can be approximated as quasi one dimensional scattering problem. The reflection in each channel or sub-sample provides two parameters: rms of the reflection spectra and the correlation of the reflection spectra. From these two quantities, the effective structural disorder or nanoscale "disorder strength $L_d$" was characterized within biological cells (briefly shown in the later section). Where $L_d = <dn^2> \times l_c$, $<dn^2>^{1/2}$ is the strength of the refractive index fluctuations related to the mass density fluctuations in the cell, and $l_c$ is the spatial correlation decay length of the spatial refractive index fluctuations[9,3,4]. In this paper, we study the effectiveness of chemotherapy drug on cancer treatment using cancer cell lines. In particular, using prostate cancer cell lines, we have shown that the intracellular structural characteristics of the chemotherapy drug action on human prostate cancer cells of different cell lines. In particular, we examined the intracellular structural properties of the drug-resistive or drug-sensitive cells, of these cell lines. The effectiveness of drug action or drug-resistance can be evaluated using the cells' intracellular structural optical disorder strength parameter $L_d$ value, this parameter acts as a potential biomarker for effectiveness of drug action characterization as will be shown in our results. Our results show that the drug resistive cancer cells have different structural disorder or disorder strength $L_d$ value than the usual cancer (control) cells. From the earlier experimental results on cancer cell studies, we already know that increased in structural disorder strength correlated with the increased in the aggressiveness or more tumorigenicity, and the decreased in disorder strength were related to the less aggressiveness or less tumorigenicity state of the cell.[3-6] Therefore, these results suggest that drug-non-resistive and drug-resistive cancer cells are different in their structural disorder. These nanoscale alterations are in turn related to the nanoscale mass density fluctuations or nanoscale refractive index fluctuations. Therefore, the strength or the degree of disorder strength, $L_d$, in terms of optical scattering can be considered as a potential biomarker, whose values can provide the effectiveness of chemotherapy drug treatment, that is to differentiate between non-drug-resistive and drug-resistive cases of cancer cells.

In the following sections, we have described the PWS experiments with prostate cancer cell lines, chemotherapy drug-non-resistive and drug-resistive cells, description of the developed PWS instruments with finer focus, and the PWS experimental results of drug-non-resistive and drug-resistive cancer cells.

*1.2 Prostate cancer and effect of chemotherapy drug*

Prostate cancer is most commonly diagnosed among men and often metastasizing at later ages. It was known that 161,360 new cases were predicted in the United States for 2016 including 26,730 cases of death[9]. It is estimated to almost 19% cause of cancer mortality in the USA. One in seven men gets prostate cancer in their lifetime. Prostate cancer is prominent in the older men, 60% man above the age 66 diagnosed with prostate cancer, 1 in 39 men in America dies in prostate cancer, and it is the 3rd leading cause of death in the United States.[9] Therefore, effective detection and treatment of prostate cancer are important, especially for seniors. Chemotherapy is used to treat metastasized prostate cancer.. However, chemotherapy is often ineffective because cells of individual patients' tumors develop chemo-resistance.[11-13] Hence, it is important to know the effectiveness of the chemotherapy drug for the treatment and assess the resistance in the prostate cancer for better affective treatment and development of the new treatment drugs.

*1.3 Prostate cancer treatment and role of chemotherapy:*

Prostate cancer treatment usually includes steps of three types: radiation, surgery, and chemotherapy. These treatment types are performed in distinct orders according to the stage and size of the malignant tumor. Usually, chemotherapy administered after surgery to remove the tumor and prostate; this therapy expects to kill the metastasized cancer cells around the prostate gland as well as in the whole body. Resistance to chemotherapy is one of the main cause of treatment failure in all types of cancer treatment, including that of prostate cancer. Development of drug resistance by cancer cells is a serious problem since it ends the remission stage and leads to disease relapse. Clinically, chemotherapy drug docetaxel (Taxotere®) is frequently used to treat the advanced stages (metastasis) of prostate cancer. However, the prostate cancer cell's resistance to this drug is a major clinical problem because it was established as the primary drug therapy for metastatic prostate cancer [13]. Docetaxel resistance is correlated with the time period over which the cancer cells were/are treated with the drug and the dosages used.[14] Researchers are studying the drug resistance and are trying to understand the physiological effect of other chemotherapeutic drugs that can bypass the resistance and restore near-normal growth and division processes for the cells remaining after chemotherapy.[15,16,17]

In this study, we will assess the docetaxel resistance through the use of in vitro human prostate cancer cell line model: C4-2, DU-145 and PC3, where the individual cell line was treated with a dose of the docetaxel up to a period of 8 months using a standard protocol. In different chemo-resistance studies, the aggressiveness and invasion of the resistive cancer cells has been observed[21,22]. In addition, it has been reported that the docetaxel resistance in a cancer cell is highly associated with genetic alterations[16]..These genetic mutations lead to the rearrangement and alteration of the most basic building blocks of the cells, such as DNAs, RNAs and Lipids, in turn produce intracellular structural changes at the nanoscale in the treated cancer cells that can be detected by the conventional microscopy imaging[20]. Here, the effect of the drug on the change of the intra-cellular nano-architecture is measured through an established parameter called degree of structural disorder or disorder strength $L_d$ ($=<dn^2>\times lc$) as defined earlier, and compared to the corresponding values for non-drug treated cancer cells. The results reveal that chemo-resistant cells show a different value of mean disorder strength $L_d$ for the all the above three cell lines compared to their corresponding wild type cancer cells. The results support the hypothesis that the prostate cancer cells acquire docetaxel chemo-resistance and ultimately showed no response towards docetaxel treatments. Here we are establishing the fact that the disorder strength $L_d$ is an efficient parameter or biomarker that can characterize usual prostate cancer cell-lines and drug-resisted cancer cells using partial wave spectroscopy PWS or spectroscopic microscopy method.

To establish the fact that $L_d$ is a potential parameter or biomarker, we first established the hierarchy of the normal and prostate cancer cell lines. Then we have treated the chemotherapy drug to establish the differences in structural disorder between the drug-non-resistance and drug-resistance cancer cells.

## 2 Method

*2.1 Cell culture and development of docetaxel chemo-resistant prostate cancer cells of the following cell lines: C4-2 (PSMA+), DU145 (PSMA−) and PC-3 (PSMA−)*

Prostate Cancer (PCa) cell lines [C4-2 (PSMA+), DU145 (PSMA−) and PC-3 (PSMA−)] were developed in cell culture facilities at the University of Tennessee Health Science Center (UTHSC), and cultured in RPMI-1640 medium containing 10% (v/v) fetal bovine serum (FBS), and 2 mM L-glutamine (Invitrogen, Carlsbad, CA) and 1% (w/v) penicillin–streptomycin (Gibco, Thermo Fisher Scientific, Grand Island, NY) at 37 °C in a humidified 5% $CO_2$–95% air atmosphere (Thermo Fisher Scientific, Waltham, USA). PCa resistant lines were generated by initial treatments with docetaxel (MP Biomedicals, Santa Ana, CA) at 1 nM (from 5 µM stock) in 75 cm2 flasks for 24-48 hours. After treatment, the surviving cells were re-seeded into new flasks and allowed to recover for 1-2 days. Cells were maintained at 1 nM till 4 treatment cycles (4 TC). Gradually the concentration of docetaxel was increased to 2.5nm (6 TC). Further, the 5 ad 10 nM of Dtxl was continued till 8 and 12 TC, respectively. Then all cells underwent 12 TC and 15 TC with Dtxl 15 ad 20 nM, respectively. Finally, at 30 nM Dtxl we have performed 25 TC, to ensure the acquisition of chemo-resistance in all C4-2, DU145, and PC-3 cells. Following each treatment, cells were allowed to fully recuperate before assessing their resistance to docetaxel and any experimental work.[3] Since the passage number of the drug treated cells increased over time, a subset of PCa cells were aged alongside these cells as an appropriate control to ensure that the effects seen were due to resistance rather than due to an ageing effect of the PCa cells.

*2.2 Cell imaging and analysis*

Frozen cell batches with similar passages were thawed and used for all experiments. For image analysis studies, we have seeded $2.5 \times 10^4$ in each well of 4- chambered slides (Sarstedt. Inc, Newton, NC) and allowed to grow. Cells after reaching 70-80% of confluence, they were fixed using 4% para-formaldehyde for 20 mins. After incubation cells were washed with PBS and cells were imaged, analyzed [17, 18, 19].

# 3 Optical partial wave spectroscopy (PWS) experiment

## 3.1 Instrumentation

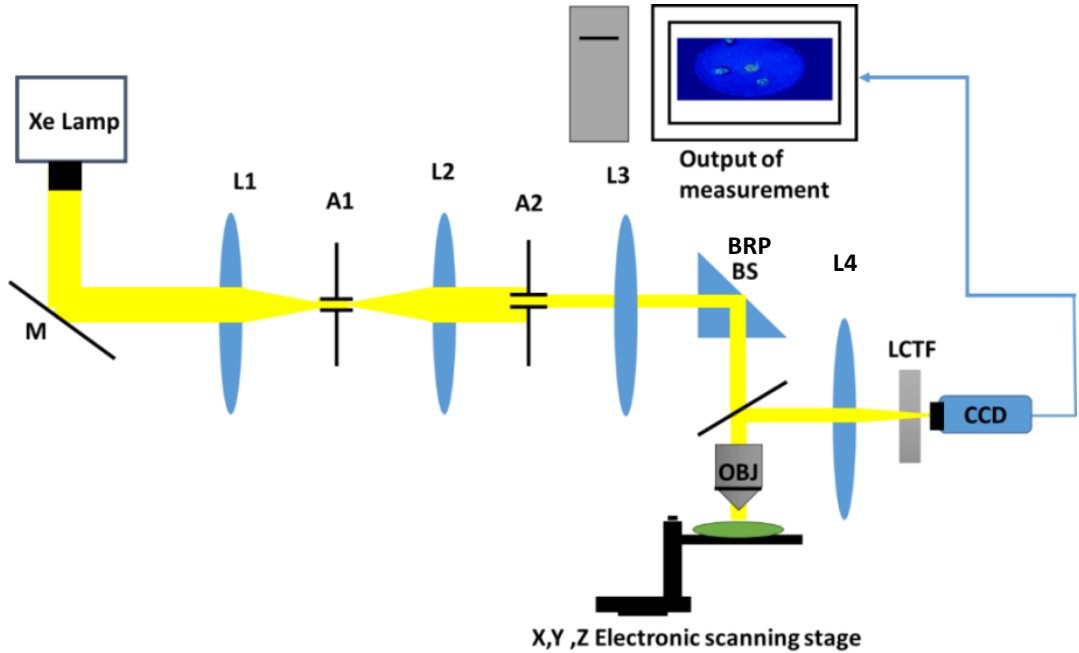

**Fig. 1** Experimental setup for partial wave microscopy spectroscopy (PWS). A broad band light is focused on a beam splitter and passes through a microscopy objective and reflected back from the sample. The backscattered light from the sample is focused onto a LCTF coupled with a CCD camera.

A schematic of experimental apparatus used for partial wave spectroscopic microscopy is shown in figure 1. This PWS setup is the version developed at the University of Memphis that provide finer focus. A broadband of white light from a Xenon Lamp (Thor Lab) is used to illuminate the sample using the technique of Kohler illumination. First the light directed through a broadband dielectric round mirror (dimeter=50mm, R>99%, thickness=12.7mm ) to pass through a set of lenses (convex, f=50mm) and apertures to collimate the light as shown in the Fig. 1. Then the light is reflected by a right angle prism (BRP, 25.mm) to let it pass through a beam splitter plate (BSP, 50:50) to a low numerical aperture objective (NA=0.65, 40x). Then, the beam focused on the sample which is kept on electronic motorized stage (Zaber Tech Inc.) with precise resolutions of 100nm vertical (z-direction) and 40 nm in x-y plane and so that the depth of the image was focused more accurately that from the previous versions[3,4]. This element provided the control on the accurate depth of focus in turn the reflection from the sample. The backscattered light passes through the objective again and is projected into a liquid crystal tunable filter (LCTF) (Verispace, LLC) where the signal is filtered according to its wavelength components. The spectral resolution of the LCTF is 1nm. The spectral range is the visible range (450-700nm).

The filtered signal is captured by the CCD camera detector (Cool Snap, LLC), which is coupled to the LCTF.

*3.2 Derivation of value of $L_d$ from backscattering intensity*

The CCD camera detect and store the backscattered image of the biological cell at every wavelength (λ) in the visible range, 450nm -700nm. Each image is represented by a matrix with (x,y) the spatial position within the matrix, and each matrix element is captured at one wavelength (λ) at a time. Eventually, we acquire a cube data from the PWS system *I(x,y;λ)* and further processing for the data, we extract the backscatter spectrum fluctuation *R( x,y λ)*. Imaged cells were quite thin (~1 µm) while lying on glass slides, therefore, the backscattered intensities could be decomposed using a simultaneous parallel backscattering channel assumption for the back reflection. In addition, the back-reflection data could be analyzed based on the mesoscopic physics framework. In this assumption $L_d$ was derived from the two quantities, the RMS value of the reflection intensity $<R>_{rms}$ and the spectral correlation decay of the reflection intensity *C(Δk)*, as described in our earlier publications.[3,4] It can be shown that:

$$L_d = \frac{B \langle R \rangle_{rms}}{2k^2} \frac{(\Delta k)^2}{-\ln(C(\Delta k))}\bigg|_{\Delta k \to 0}$$

where *B* is the normalization constant and $C(\Delta k)$ is the autocorrelation function of *R(x,y,k)* at a particular position (x,y) and averaged over many ensembles. The acquisition time for each cell is ~1 min. The data were collected, and the average of $L_d$ was calculated for each point (x,y). It can be further shown that.

$$L_d \sim <dn^2> \times l_c.$$

where $<dn^2>$ is the variance of the refractive index and $l_c$ is the spatial correlation length of the refractivity index fluctuation of an optical sample.3,4

## 4  Results and Discussions

To demonstrate the ability of the present PWS instrument to identify nano-architectural changes in cells with different levels of tumorigenicity and modifications of these tumorigenicity level due to the induced drug resistance of the cells by drug exposure, we performed our first set of experiments on different prostate human adenocarcinoma cell lines. We focused on the prostate cancer given that it is a major public health problem in the United States. Moreover, chemotherapy-induced drug resistance is an observable clinical problem. We chose three prostate cancer cell lines with already known extents of tumorigenicity, low, moderate, and very high, which were, DU145, C4-2 and PC3 respectively.  PWS structural disorder measurements were conducted on ~ 20 cells randomly selected from each cell type for one single measurement (~60 cells for all 3 cells types). Subsequently, mesoscopic physics based analysis of the backscattered data matrix *R(λ, x, y)* was performed on the sets of data for each cell type.

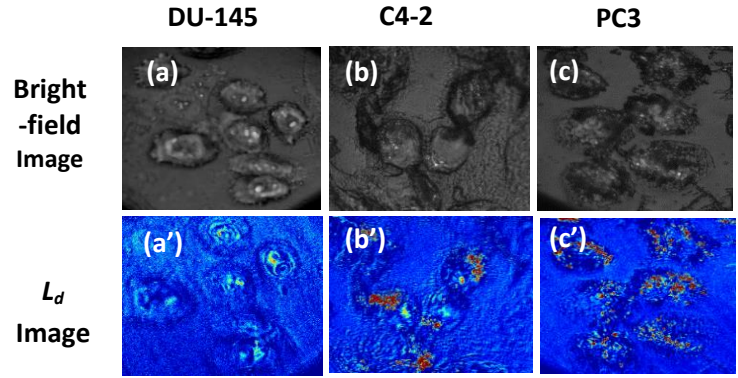

**Fig. 2** Representative bright field images (a,b,c) and corresponding 2D $L_d$ maps (a',b',c') (PWS images) for the three prostate cell line DU145, C4-2, PC3,1,2,3 respectively.

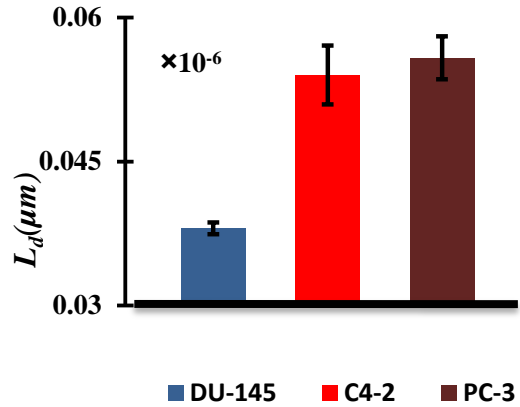

**Fig. 3** The bar plots of mean intracellular disorder strength $L_d$ calculated for three types of human prostate cancer cells (DU145, C4-2, PC3). The $L_d$ results show that the average $L_d$ value correlate with the tumorigenicity level of the cell.

In Fig. 2, we have extracted the reflectance part $R(\lambda,x,y)$ and its correlation length to calculate the disorder strength $L_d$ parameter for each pixel of CCD in the captured PWS images, as described in the method. The $L_d$ value was calculated for each pixel in the individual cell image and represented by plot of 2D color maps, and named as an PWS $L_d$ map. In Figs. 2 (a)-(c), three representative bright field image of each cell line is presented, and the corresponding PWS $L_d$ maps are shown in Figs. 2(a')-(c'). The bright filed images, as shown in (a), (b), and (c) of the three types of cells appear not differentiable, however corresponding $L_d$ maps, as shown in figures (a'), (b'), and (c') that are clearly different. The red spots are corresponding to the higher disorder strength, of higher $L_d$ value of that pixel.

For statistical comparison, the corresponding bars in graph of Fig. 3 shows the average $L_d$ from 20 cells from each type of cell line and the corresponding error bars represent the standard error of the mean. In Fig. 3, significant statistical differences in the local disorder strength $L_d$ values are visually obvious according to the number of red regions displayed for the three-distinct prostate cancer cell lines. Clearly, the fraction of higher $L_d$ values increased in the more aggressive cancer cells. Importantly, the most aggressive cell line (PC3) has the highest intercellular disorder strength, while the least aggressive cell line (Du145) exhibited the least structural disorder or disorder strength. These results suggest higher disorder strength is associated with an increase in the tumorigenicity level in the cell.

In the next step, we assessed the structural effects that may introduced on the prostate cancer cells from the three-cell lines due to the prolonged, hopefully therapeutic treatment with the chemotherapy drug docetaxel. We hypothesized that prostate cancer cell lines may develop some type of resistance to the drug after a prolong time of treatment (8 months in this case), and that this behavior would be associated with additional intracellular structural changes as well.

In particular, an imprint of structural disorder in the drug resistive prostate cancer cells was expected. To test our hypothesis, we performed PWS measurement on docetaxel treated cells of cell lines PC3, C4-2, and DU145, as well as on corresponding age matched cell lines. These tests may answer the question if the drug-resistant cells showed measurable structural changes associated with the aggressiveness behavior that prostate cancer cells acquired after 8 months of docetaxel chemotherapy drug treatment.

Using image collection and value extraction as described above, we first observed the disorder strength for the docetaxol-treated cells (resistance cells are noted by R) from each type of prostate cancer cell line, and compared the $L_d$ parameter calculated for the drug treated cells with the $L_d$ calculated for cancer cells in the first set of the experiment by considering the later as wild (non-resistive to drug) cancer cells. In Fig. 4, we have plotted the 2D $L_d$ maps for the three representative resistance cell lines and for the corresponding control (cancer) cells. The maps use a color bar from blue to red to show an increasing value of the disorder strength $L_d$ value. Clear differences are visible in the PWS $L_d$ maps for the control and the resistance cell for each type of cells individually representing the three types of cells lines; whereas, in contrast to that display, the bright field images of these cells types look quite similar and definite features are difficult to quantify. Most interestingly, the 2D maps of $L_d$ of the resistance cells showed higher values of $L_d$ compared to the control cells and they were consistent for the three type of the prostate cancer cell lines.

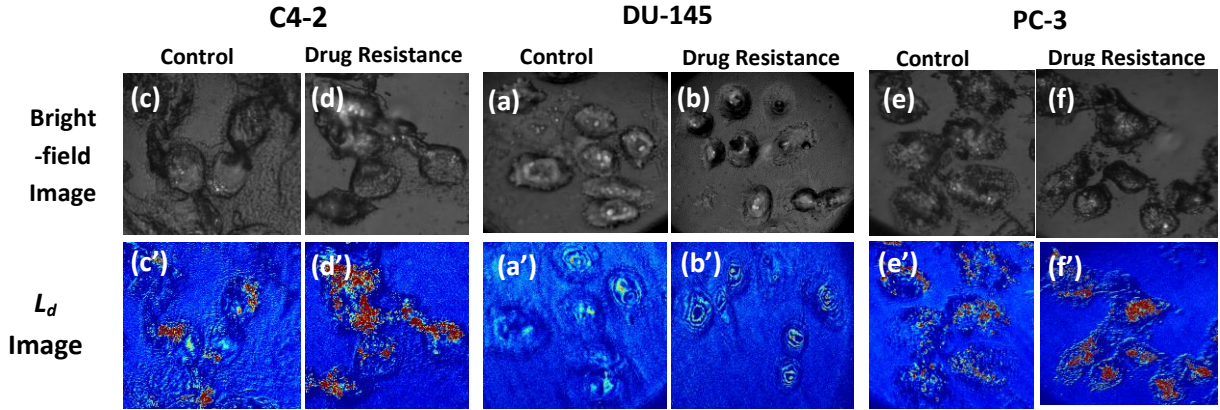

**Fig. 4** (a-f) Bright field images of three types of human prostate cell line: (control) and its corresponding (resistance R) cells from the same cell type. (a'-f'). The colored images are the PWS images, a 2D map of $L_d$, images show the representative mean intracellular disorder strength between the control and the corresponding drug resistance prostate cells.

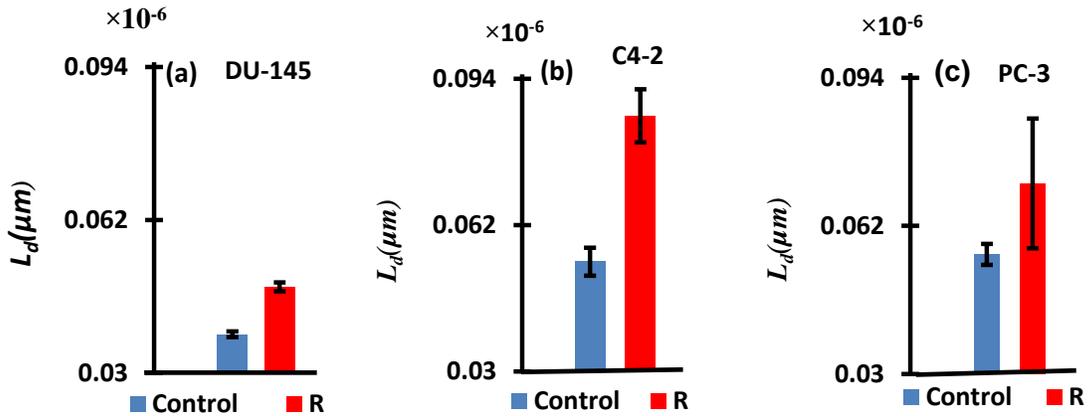

**Fig. 5** Bar plots for mean intracellular disorder strength value $L_d$ calculated for three types of human prostate cancer cells from cell lines: (a) C3, (b) C4-2, and (c) DU-145) (control vector), and compared with the mean intercellular disorder strength of the corresponding cancer cells treated with docetaxel drug for about 8 months. The drug resistive survival cells are called drug resistance cells ( noted by R : drug resistance). It can be seen that the each drug resistance cell type has higher $L_d$ than its corresponding control/wild type.

To further show that the structural differences between the control and the drug-resistance cells, in Fig. 5, we plotted the bar graphs for mean $L_d$ values for the control and the resistant cells averaged over 20 cells for each cell type. These data are shown in Fig.5, where the mean $L_d$ value is significantly elevated in the drug-resistant cells that were induced with docetaxel, as compared with the untreated cells.

The results not only show the cancer cells underwent changes in their nanoarchitecture associated with the prolonged exposure to the chemotherapy drug, but also show the potential use and quantitative capacity of the disorder strength, $L_d$, as a parameter or biomarker for accessing structural changes measured at the nanoscale in the cancerous cells, that are associated with drug-exposure resistance.

## 5 Conclusions

In summary, we have shown the application of partial wave spectroscopy as developed in the present apparatus to demonstrate readily detectable differences in cells at selected stages of prostate cancer. The stages were selected because they represent levels of tumorigenicity for prostate cancer cells, and accordingly the measurements are likely to have clinical utility. Human prostate cancer cell line samples, the control cells (or non-drug-exposed cells) and corresponding cells exhibiting drug-resistance linked to prior exposure to the chemotherapy drug docetaxel, showed a significant difference in their degree of structural disorder strength $L_d$ values. Cancer cells were treated with docetaxel, the mean disorder strength of the cells that survived drug exposure was higher as shown by $L_d$ values compare to the $L_d$ value of the control (docetaxel chemo-sensitive cells) samples for all three cell types: PC3, C4-2 and DU-145.

The results suggest that there are different structural alterations have happened in the structures of the prostate cells for control (non-drug-resistance cancer cells) and drug-resistive cancer cells. These alterations presented as a structural alteration or mass-density fluctuations increases in survival cells when the drug is introduced and used for a long period.

The drug-resistant cells had greater ranges of structure than the drug-sensitive cells and this is reflected in their intracellular structural disorder properties, or disorder strength. There are different mechanism for cancerous cells going for drug resistance. Such as: drug inactivation, alteration of drug targets, DNA damage repair, cell death inhibition, epithelial-mesenchymal transition and metastasis, cancer cell heterogeneity, cancer cell heterogeneity etc,.[23-26] As the drug resistive cells are surviving more due to the different pathways of the drug resistance to the cancer cells as few are listed above, the cancer cells increase their aggressiveness, in turn increase in the structural disorder due to random growth. And this extra increase in the structural disorder in drug resistive cancer cells can be measured and shown to be a potential biomarker to access the effect of chemotherapy drug.

These experimental results provided a new insight to the chemotherapy drug-resistance cells and associated increased in the structural disorder. The origin of these extra structural changes and their correlations with the specific molecular changes, and importantly how these structural changes are related to the drug resistance, are yet to be explored in details. Yet, when they are, the local value of $L_d$ can be matched to the submicroscopic details.

**Acknowledgments**

National Institutes of Health (NIH) grants (Nos. R01EB003682 and R01EB016983); Faculty Research Award from the University of Memphis; FedEx Institute of Technlology grant, for Pradhan. Dr. Yallapu was supported by NIH K22 CA1748841, NIH R15 CA213232, UTHSC CORNET Grants.